\documentclass[11pt]{article}

\usepackage[utf8]{inputenc}
\usepackage[T1]{fontenc}
\usepackage{times} % Usa fontes Times
\usepackage{amsmath, amssymb, amsfonts}
\usepackage{graphicx}
\usepackage[margin=1in]{geometry}
\usepackage{authblk} % Para afiliação de autor
\usepackage{enumitem}
\usepackage{microtype} % Melhora o espaçamento do texto
\usepackage{booktabs}
\usepackage{longtable}
\usepackage{caption}
\usepackage{array}

\usepackage{xcolor}
\usepackage{textcomp}     % Para símbolos como a seta
\usepackage{xurl}         % Para quebra de URL robusta
\usepackage{listings}     % Para formatar código

\usepackage{hyperref}
\hypersetup{
    colorlinks=true,
    linkcolor=blue,
    filecolor=magenta,
    urlcolor=cyan,
    citecolor=blue
}

\lstdefinestyle{json}{
    language=json,
    basicstyle=\ttfamily\scriptsize, % Tamanho de fonte um pouco maior
    numbers=left,
    numberstyle=\tiny\color{gray},
    stepnumber=1,
    numbersep=5pt,
    showstringspaces=false,
    breaklines=true,
    frame=tb, % Apenas linhas em cima e em baixo (top, bottom)
    backgroundcolor=\color{gray!5},
    stringstyle=\color{purple!80!black},
    keywordstyle=\color{blue!70!black},
    commentstyle=\color{green!60!black},
    morestring=[b]",
    escapechar=|, % Habilita o caractere de escape '|'
    literate=
      {\{}{{{\color{blue!70!black}\lbrace}}}1
      {\}}{{{\color{blue!70!black}\rbrace}}}1
      {[}{{{\color{blue!70!black}[}}}1
      {]}{{{\color{blue!70!black}]}}}1,
}

\lstdefinestyle{tablecode}{
  basicstyle=\small\ttfamily,
  breaklines=true,
  breakatwhitespace=false,
  postbreak=\mbox{\textcolor{red}{\textrightarrow}\space},
  frame=none,
  showstringspaces=false,
  tabsize=2,
  columns=fullflexible, % Permite que o texto flua melhor
  literate=
    *{_}{{\_}}1
     {.}{.}{1}
     {;}{;}{1}
     {:}{:}{1}
     {/}{/}{1}
     {?}{?}{1}
     {=}{=}{1},
}

\title{Legal Knowledge Graph Foundations, Part I: \\ URI-Addressable Abstract Works (LRMoo F1 → schema.org)}

\author[1]{Hudson de Martim}
\affil[1]{Federal Senate of Brazil \\ \texttt{hudsonm@senado.leg.br}}
\date{}

\begin{document}

\maketitle

\begin{abstract}
Building upon a formal, event-centric model for the diachronic evolution of legal norms grounded in the IFLA Library Reference Model (LRMoo) \cite{demartim2025temporal}, this paper addresses the essential first step of publishing this model's foundational entity—the abstract legal Work (F1)—on the Semantic Web. We propose a detailed, property-by-property mapping of the LRMoo F1 Work to the widely adopted schema.org/Legislation vocabulary. Using Brazilian federal legislation from the Normas.leg.br portal as a practical case study, we demonstrate how to create interoperable, machine-readable descriptions via JSON-LD, focusing on stable URN identifiers, core metadata, and norm relationships. This structured mapping establishes a stable, URI-addressable anchor for each legal norm, creating a verifiable "ground truth". It provides the essential, interoperable foundation upon which subsequent layers of the model, such as temporal versions (Expressions) and internal components, can be built. By bridging formal ontology with web-native standards, this work paves the way for building deterministic and reliable Legal Knowledge Graphs (LKGs), overcoming the limitations of purely probabilistic models.
\end{abstract}


\section{Introduction}

Open Government Data (OGD) initiatives are fundamental for enhancing transparency and public participation by making government-generated information openly accessible. However, while the principles of OGD are well established, their application to the legal domain presents unique and significant challenges. Legal norms are not static data points. They are complex, structured documents with a dynamic lifecycle, characterized by a formal hierarchy, extensive inter-norm references, and continuous evolution through temporal versions.

Simply publishing legal texts as open files (e.g., PDF) fails to capture this rich semantic structure. This hinders the development of advanced applications that require machine-readable data. Building upon a formal, LRMoo-based ontological model for the diachronic evolution of legal norms \cite{demartim2025temporal}, this paper presents the first foundational step in translating that formal model to the web. We detail a structured, standardized mapping for the abstract legal Work (F1)—the conceptual essence of a legal norm—using the widely adopted Schema.org vocabulary.

\subsection{The Significance of Open Government Data}
Open Government Data (OGD) initiatives are pivotal for enhancing transparency, accountability, and innovation by publishing public-sector information without restrictions \cite{ubaldi2013ogd}. By making data produced with public funds freely available, governments can stimulate economic growth, improve internal workflows, and empower citizens to participate in policymaking \cite{lathrop2010open}. While this principle of openness is fundamental for the legal domain, its effective implementation demands more than simple data availability; it requires addressing the unique structural and temporal complexities of legal norms.

\subsection{The Challenge of Structuring Legal Norms for Machine Readability}
While OGD principles provide the "why," the nature of legal data dictates the "how." Unlike tabular data, legal norms possess a deep, hierarchical structure and evolve over time through amendments and repeals. This temporal dynamism means that the "correct" text of a law is dependent on a specific point in time, a challenge formally addressed by event-centric ontologies \cite{demartim2025temporal}. The dense web of citations between norms further creates a complex graph of relationships lost in unstructured formats.

This lack of structured data remains a critical bottleneck for Legal Tech. Advanced systems require a deterministic, verifiable "ground truth" of the law, which cannot be achieved without a formal data model. Such a foundation is a prerequisite for reliable AI-powered retrieval systems, including sophisticated approaches like the ontology-driven SAT Graph RAG framework \cite{demartim2025graphrag}.

\subsection{Our Contribution: A Schema.org Mapping for Brazilian Legal Works}
To address these challenges, this paper proposes a unified mapping of Brazilian legislation to the \url{schema.org/Legislation} vocabulary. Our approach centers on the foundational concept of the legal Work, as defined by the IFLA Library Reference Model (LRM).
This paper's core contribution is the detailed specification for mapping the abstract legal Work (F1) to \texttt{sdo:Legislation}. As established in our foundational ontological model \cite{demartim2025temporal}, the Work represents the legal instrument as a persistent intellectual creation, independent of its specific textual versions or formats. We detail the key properties for describing this entity via JSON-LD, providing concrete examples from the Brazilian Normas.leg.br portal.
By defining a stable, URI-addressable anchor for each legal norm, this work establishes the indispensable foundation for modeling the norm's full lifecycle. This includes its internal components (Component Works) and its dynamic temporal and linguistic versions (Expressions), as formally structured in our LRMoo-based framework \cite{demartim2025temporal}.

\section{Background and Related Works}
The effective use of legal norms in automated systems demands a shift from plain text to structured, machine-readable formats. This transition is the bedrock for the modern Legal Tech sector, enabling tools that manage and analyze legal information at scale  \cite{veith2016legaltech}. Legal Knowledge Graphs (LKGs) in particular—which represent legal entities and their relationships as graph structures—are fundamentally dependent on such semantically rich data to support sophisticated queries and automated inference \cite{filtz2017building, koniaris2016structuring}.

This section outlines the foundational standards and principles our work is built upon. We ground our approach in two key pillars: a technical framework comprising the cornerstones of Linked Data and the Schema.org vocabulary, and a conceptual framework based on the IFLA Library Reference Model (LRMoo). We then review related standards for legal data modeling to contextualize our contribution.

\subsection{Foundational Technologies}
\subsubsection{The Linked Data Principles}
The publication of structured data on the web is guided by the four core principles of Linked Data, which aim to create a global, machine-readable web of interconnected information \cite{bizer2008linkeddata}:
\begin{enumerate}
    \item \textbf{Use URIs to name everything.} Assign a unique URI to each legal norm, ensuring unambiguous identification.
    \item \textbf{Use HTTP(S) URIs.} Publish these identifiers so they can be accessed and managed over the Web.
    \item \textbf{Make URIs dereferenceable.} Configure each URI to return machine-readable metadata about the identified resource.
    \item \textbf{Include links to other URIs.} Create relationships between legal entities to form a web of interconnected data.
\end{enumerate}

These principles ensure clarity in identifier management and lay the groundwork for the seamless integration of any structured data, such as legal information, into a global, interconnected knowledge graph.

\subsubsection{Schema.org Vocabulary: A Bridge for the LRMoo Work Entity} 
To describe the resources identified by URIs, we use \textbf{Schema.org}\footnote{Schema.org is a collaborative project providing a shared vocabulary for structured data markup on the internet, primarily used to help search engines understand web content. The project Portal can be accessed at \url{https://schema.org/}}, a shared, collaborative vocabulary for structured data. We leverage its hierarchical structure, starting from the generic \textbf{sdo:Thing} type.

While the \textbf{sdo:CreativeWork} type serves as a general parent for creative outputs, Schema.org provides the more specific and semantically rich subclass: \textbf{sdo:Legislation}. This type is the ideal choice for representing the abstract F1 Work entity from our foundational LRMoo model \cite{demartim2025temporal}. The alignment is evident in its data properties: properties such as legislationIdentifier, legislationDate (enactment), and jurisdiction directly correspond to the persistent, high-level metadata that define a legal norm's conceptual identity. These attributes remain constant across all its subsequent temporal versions (Expressions), making sdo:Legislation the perfect vehicle for modeling the stable anchor of a legal norm's lifecycle.

Furthermore, this suitability is not coincidental. The schema.org/Legislation vocabulary was heavily inspired by the mature European Legislation Identifier (ELI) ontology \cite{eli-guide}, a standard itself grounded in the principles of the FRBR/LRMoo conceptual model. This strong lineage underscores the fitness of sdo:Legislation for building a standardized and interoperable bridge between formal bibliographic models and the discoverability of the open web.

It is important, however, to acknowledge the nature of the Schema.org vocabulary. Unlike formal ontologies such as CIDOC CRM and LRMoo, which prioritize conceptual precision through a detailed, event-centric model, Schema.org is designed for practical, large-scale web data annotation. Consequently, it employs numerous "shortcuts" or flattened properties. For instance, properties like \texttt{legislationDate} and \texttt{publisher} are attached directly to the \texttt{Legislation} type, whereas a formal ontology would model these as attributes of distinct \textit{Creation} and \textit{Publication} events.
Despite this ontological imprecision, Schema.org's utility for describing the metadata of a Legal Knowledge Graph is immense. Its value lies in three key areas:
\begin{itemize}[noitemsep, topsep=0pt]
\item \textbf{Interoperability and Discoverability:} As the de facto standard for structured data on the web, it makes legal metadata immediately understandable to a vast ecosystem of tools, from major search engines like Google to generic data crawlers. This ensures that the LKG's foundational entities are not isolated in a specialized silo but are first-class citizens of the global web of data.
\item \textbf{Low Barrier to Entry:} Its simpler, attribute-based structure makes it easier for data publishers to implement, fostering wider adoption compared to more complex formal ontologies. This pragmatic approach is fundamental for scaling up Open Government Data initiatives.
\item \textbf{A Bridge to Deeper Models:} A well-structured Schema.org description can serve as a high-level "entry point" or an interoperable metadata layer for a much richer, LRMoo-aligned backend. The simplified properties can be automatically expanded into their full event-centric representations within the LKG, offering both web-native simplicity on the surface and ontological rigor in the core.
\end{itemize}
Therefore, our choice of Schema.org is a strategic one: we leverage its unparalleled reach and simplicity as a foundational layer for discovery, while grounding our core model in the formal precision of LRMoo, as detailed in the subsequent mapping.

\begin{figure}[htbp]
    \centering
    \includegraphics[width=1.0\textwidth]{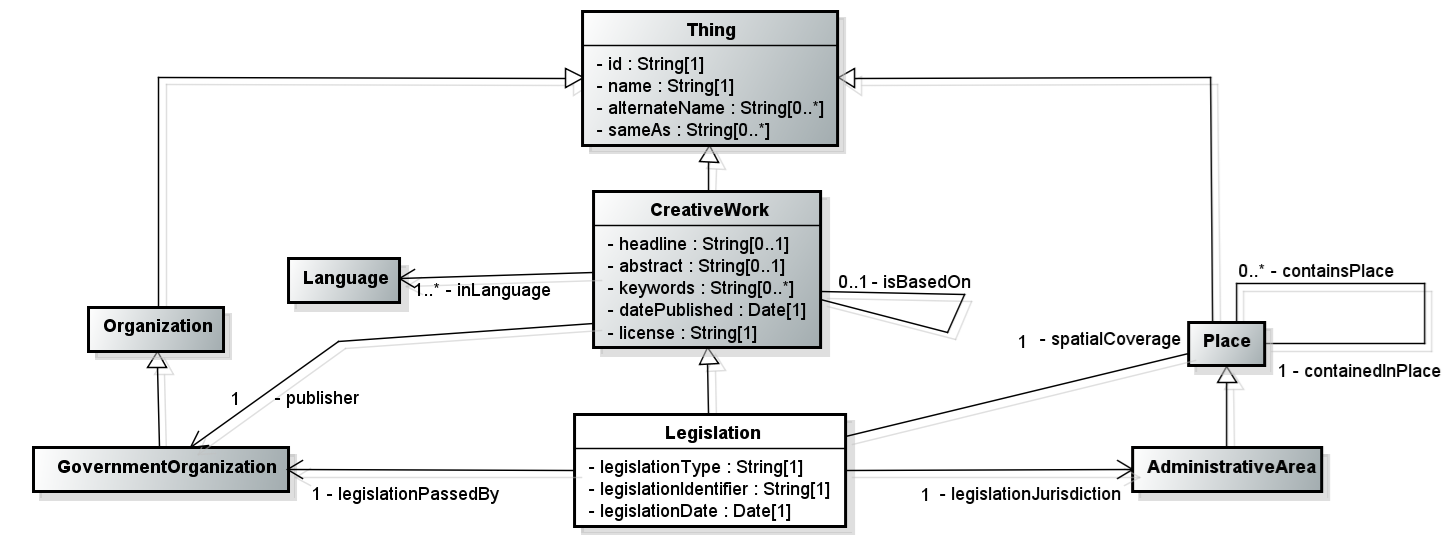} 
    \caption{Simplified diagram of the Schema.org vocabulary, according to the proposed mapping in this work.}
    \label{fig:schema_diag1}
\end{figure}

Figure \ref{fig:schema_diag1} provides a simplified diagram of this vocabulary, focusing on the subset of types and properties used in this paper to model the legal Work. Notably, the relationship cardinalities have been defined to reflect their concrete application within this foundational layer of our proposed LKG schema.

\subsection{Related Works}
The challenge of representing legal norms in a machine-readable format has been addressed by several mature standards, each with a distinct focus. Our work positions itself within this landscape by intentionally focusing on a different layer of abstraction: instead of encoding the deep internal structure of documents, we prioritize the publication of high-level, web-native metadata that makes the legal norm a discoverable and linkable entity.

\subsubsection{Akoma Ntoso (AKN)}

A prominent standard for encoding the internal structure of legal documents is \textbf{Akoma Ntoso (AKN)} \cite{palmirani2011akoma}. As an XML schema, AKN provides a rich vocabulary for marking up the hierarchical components of a legal text (e.g., articles, paragraphs) and identifying textual modifications. AKN's metadata block explicitly adopts the FRBR model to distinguish between the abstract \texttt{Work}, its versioned \texttt{Expressions}, and formatted \texttt{Manifestations}. While AKN offers a powerful solution for document-centric representation, its primary focus remains on the granular encoding of the document itself. Its XML-based format is ideal for creating a definitive, machine-readable version of the legal text for archival and processing pipelines. Our approach is therefore complementary: while AKN provides the definitive, machine-readable content for archival and processing pipelines, our proposal provides the essential context for its discovery and integration into a global knowledge graph.

\subsubsection{LexML Brazil}

In the Brazilian context, the \textbf{LexML Brazil} project provides a foundational framework for legal information, offering both an XML schema for document representation, inspired by Akoma Ntoso, and a standardized URN syntax for uniquely identifying norms \cite{lexml2008modelo}. This URN scheme is particularly powerful, as its syntax allows for the precise addressing of specific versions, language editions, and even internal components of a legal norm.

Our work builds upon and enriches this robust identification system. While a URN serves as a persistent and unambiguous pointer to a resource, it is designed for identification, not description. This distinction is a core tenet of our foundational modeling work \cite{demartim2025temporal}, which argues that a separate layer of semantic metadata is needed to describe what the resource is and how it relates to other entities. The mapping proposed in this paper provides precisely this semantic layer. We address this by using the LexML URN as the value for the \texttt{legislationIdentifier} property within a rich, structured schema.org/Legislation description. In doing so, we attach discoverable, web-native semantics to the powerful identifiers that LexML already provides, transforming a simple pointer into a fully described entity within the web of data.

\subsubsection{European Legislation Identifier (ELI)}

Our work's alignment with the principles of the \textbf{European Legislation Identifier (ELI)} \cite{eli-guide} is a direct consequence of our choice of vocabulary. As established in Section 2.1.2, the \texttt{schema.org/Legislation} model inherits its design from the ELI ontology. Therefore, our mapping is not merely compatible with ELI but can be seen as a direct implementation of its core principles—such as persistent identification and structured metadata—for the Brazilian legal context. A key contribution of our approach is the operationalization of these specialized principles within a general-purpose, globally recognized vocabulary. By leveraging the reach of Schema.org, we aim to lower the barrier for data integration and ensure that legal metadata becomes a first-class citizen on the open web, immediately consumable by search engines and other generic data crawlers.

\section{Mapping the Legal Work to \texttt{sdo:Legislation}} 
\label{sec:mapping_work}
The cornerstone of a robust Legal Knowledge Graph is a formal model that maintains the identity of legal norms through time and across formats. To achieve this, we adopt the conceptual framework of the IFLA Library Reference Model (LRMoo), the official ontological model of LRM and successor to FRBRoo \cite{lrmoo2024}. As detailed in our foundational work on modeling legal evolution \cite{demartim2025temporal}, this event-centric model distinguishes between the abstract intellectual content and its various realizations.

Following this model, we distinguish between three core entities. First, an abstract F1 Work is a distinct intellectual creation—the law's conceptual essence (e.g., the "1988 Brazilian Federal Constitution")—which persists through time, independent of any specific wording. This Work is realized through one or more F2 Expressions, which are the specific, versioned textual contents of the law (e.g., its original 1988 text or its text as of 2025). Finally, an Expression is embodied in a F3 Manifestation, the physical or digital format in which it is published (e.g., a PDF file or an HTML web page). This paper, as the first part of our proposed framework, focuses exclusively on the foundational F1 Work, which serves as the stable anchor for the entire lifecycle of a legal norm.

To represent this abstract Work using a widely-understood, interoperable vocabulary, we turn to Schema.org. While a legal norm could be typed as a generic sdo:CreativeWork, Schema.org provides the more specific and semantically rich subclass: sdo:Legislation. The definition of sdo:Legislation as "a legal act, enforceable or not..." aligns directly with our conceptual Work entity. This alignment is confirmed by its data properties: properties such as sdo:jurisdiction, sdo:legislationType, and sdo:legislationIdentifier correspond directly to the persistent, high-level metadata of the abstract legal Work, which remain constant across all its temporal versions (Expressions). This strong alignment in both definition and data structure makes sdo:Legislation the ideal choice for modeling the top-level Work entity in our framework.

\section{Case Study: Implementation for Brazilian Federal Legislation}
To demonstrate and validate our proposed mapping, we apply it to the context of Brazilian federal legislation, using the \textbf{Normas.leg.br} portal as our practical case study. Maintained by the Brazilian National Congress, Normas.leg.br serves as an implementation of Open Government Data principles for the legal domain, providing structured access to federal norms' texts and metadata\footnote{The Normas.leg.br portal can be accessed at: \url{https://normas.leg.br}. Last accessed on 2025-04-19.}. The project strategically adopted the principles of Linked Data and the Schema.org vocabulary to make Brazilian legislation discoverable and integrable into a global Legal Knowledge Graph \cite{demartim2019proposal, bizer2008ldow}.

This case study details the property-by-property mapping for representing a Brazilian legal \textit{Work} as a \texttt{sdo:Legislation} entity. Before presenting the detailed mapping table and a concrete JSON-LD example, it is helpful to visualize the underlying event-centric model from LRMoo. This clarifies how properties like \texttt{legislationDate} or \texttt{publisher} in Schema.org act as convenient shortcuts for describing complex creation and publication events.

\subsection{An Event-Centric View of the Legal Lifecycle}

The lifecycle of a legal norm, from its signature to its publication, can be modeled as a chain of discrete events. Each event has participants, a time-span, and produces specific outputs.

\subsubsection{The Enactment Event}
\begin{figure}[htbp]
    \centering
    \includegraphics[width=0.9\textwidth]{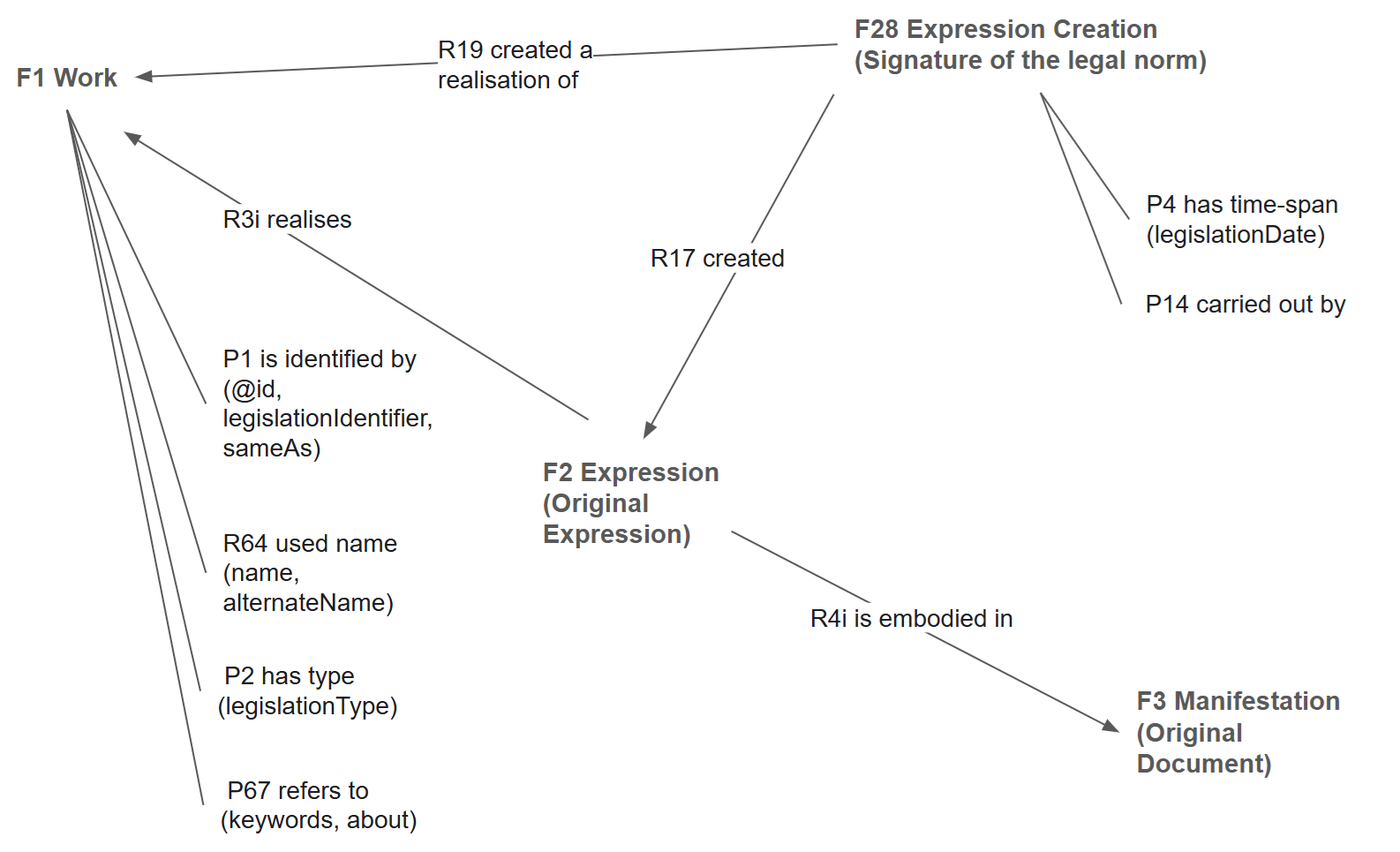} 
    \caption{LRMoo model of the enactment event (\texttt{F28 Expression Creation}).}
    \label{fig:enactment_event}
\end{figure}

The creation of a legal norm is modeled as an \texttt{F28 Expression Creation} event, as shown in Figure~\ref{fig:enactment_event}. This single event represents the act of signature or promulgation and has two closely-related outcomes: it creates (\texttt{R17 created}) the \texttt{F2 Expression} — the textual/linguistic instantiation of the norm — and (coinciding with that act) results in the first \texttt{F3 Manifestation} that \texttt{R4 embodies} the newly created expression (i.e., the physical or digital original document that was signed). The event also formally establishes that the new expression realises the abstract \texttt{F1 Work} (\texttt{R19 created a realisation of}). 

The property \texttt{legislationDate} from Schema.org, presented in \ref{sec:properties_mapping}, is not an intrinsic attribute of the abstract \texttt{F1 Work}. Instead, it is a convenient shortcut that semantically derive from this specific \texttt{F28} enactment event (date of enactment).

\subsubsection{The Publication Process}

The publication of a legal norm is rarely an isolated event. In many legal traditions, including Brazil, a norm becomes legally binding when it is incorporated into a larger official publication, such as the Official Gazette ("Diário Oficial" in Brazil). To represent this reality with high fidelity, we adopt an adaptation to LRMoo of the model proposed by the LexML Brazil project \cite{lexml2008modelo}.

In this model, the specific edition of the Gazette is itself a distinct work. Its complete textual expression, created by an \texttt{F28 Expression Creation} event, makes the crucial connection: it incorporates the individual expression of the legal norm. A subsequent \textbf{publication event} then produces the final, public document (a \texttt{F3 Manifestation}) which materializes the entire Gazette's expression, including the embedded norm.

This event-centric view has a key practical consequence. Properties commonly attached to a norm's metadata, such as \texttt{datePublished} and \texttt{publisher}, presented in \ref{sec:properties_mapping}, are not intrinsic attributes of the abstract \texttt{F1 Work}. Instead, they are convenient shortcuts that semantically derive from the publication event of the containing Gazette edition. While exposing these shortcuts directly on a \texttt{schema:Legislation} resource is acceptable for ease of consumption, systems requiring authoritative provenance should trace the relationships from the Work to the publication event and its agent. This enables precise queries like "Which Gazette issue carried the legally binding text?"

\subsection{Property-by-Property Mapping}
\label{sec:properties_mapping}

This case study details the property-by-property mapping for representing a Brazilian legal \textit{Work} as a \texttt{sdo:Legislation} entity---this mapping has been available as JSON-LD annotations in the Norma.leg.br portal pages since 2019. Table \ref{tab:norm_properties} presents the key properties for describing a \texttt{sdo:Legislation} instance when it represents a legal \emph{F1 Work}. We detail the mapping for Brazilian legal norms using “Lei Complementar nº 123 de 14/12/2006” (Complementary Law No. 123 of 2006-12-14)\footnote{The general metadata of the Brazilian legal norm "Lei Complementar nº 123 de 14/12/2006" (Complementary Law No. 123 of 2006-12-14) can be visualized at \url{https://normas.leg.br/?urn=urn:lex:br:federal:lei.complementar:2006-12-14;123}. The general metadata are presented in a structured format in the information panel (on the right side of the page). Last accessed on 2025-04-19.} as a running example.

\begin{longtable}{p{0.25\linewidth} p{0.35\linewidth} p{0.35\linewidth}} 
\caption{Properties of a \texttt{sdo:Legislation} instance representing a legal \textit{F1 Work}, detailing their corresponding LRMoo/CIDOC CRM mappings, descriptions for Brazilian norms, and examples from "Lei Complementar nº 123 de 14/12/2006".}
\label{tab:norm_properties} \\
\toprule
\textbf{Properties} & \textbf{Mapping} & \textbf{Examples} \\
\midrule
\endfirsthead

\url{id} \newline (\texttt{P1 is identified by}) & The dereferenceable URI identifying the legal \emph{Work} (\texttt{E42 Identifier}). \newline It is based on the format \newline \url{https://normas.leg.br/?urn=<legislationIdentifier>} \newline The \emph{<legislationIdentifier>} property is presented below. & "\url{https://normas.leg.br/?urn=urn:lex:br:federal:lei.complementar:2006-12-14;123}" \\
\midrule

\url{legislationType} \newline (\texttt{P2 has type}) & The formal category (\texttt{E55 Type}) of the \emph{Work}\footnote{The set of possible values for types of Brazilian legislation can be found in the CategoryCodeSet declaration located at \url{https://normas.leg.br/?urn=urn:lex:br:federal:conceito:tipo.norma}. Last accessed on 2025-04-19.}. & A CategoryCode:\newline  \texttt{\newline 
\{ \newline
    "@type": "CategoryCode",
    "@id": "\url{https://normas.leg.br/?urn=urn:lex:br:federal:tipo.norma:lei.complementar}" 
\newline \}
}\\
\midrule

\url{legislationIdentifier} \newline (\texttt{P1 is identified by}) & The canonical URN of the legal \emph{Work} (\texttt{E42 Identifier}), according to the format proposed by the LexML project -- found at \url{https://projeto.lexml.gov.br/} & "\url{urn:lex:br:federal:lei.complementar:2006-12-14;123}" \\
\midrule

\url{sameAs} \newline (\texttt{P1 is identified by}) & List of alternative URLs (\texttt{E42 Identifier}) for the legal \emph{Work}. Optional. \newline For example, the URL of the norm for the LexML portal or for the portal of the associated legislative/executive House can be included. & \texttt{[ "\url{https://www.lexml.gov.br/urn/urn:lex:br:federal:lei.complementar:2006-12-14;123}", "\url{https://legis.senado.leg.br/legislacao/DetalhaSigen.action?id=572878}" ]} \\
\midrule

\url{inLanguage} & Code of the language, according to the IETF BCP 47 standard, in which the legal \emph{Work}'s metadata is described (for its content, it will be described in its \texttt{F2 Expressions})--- In LRMoo, the F1 Work is language-agnostic. The language of the content is a property of the F2 Expression (\texttt{P72 has language} (\texttt{E56 Language})).\newline If names or descriptions are provided in other languages, these should also be included. & A simple value or a list of languages: \newline \texttt{"pt"} \newline Or \newline \texttt{["pt", "en"]} \\
\midrule

\url{name} \newline (\texttt{R64 has name}) & The formal title (epigraph) of the legal \emph{Work} (\texttt{F52 Name}). \newline Populate using upper and lower case, even if the epigraph appears in all uppercase in the original text, including changing 'Nº' to 'nº'. & A simple value or a list of language-localized values: \newline "\texttt{Lei Complementar nº 123 de 14/12/2006}" \newline Or \newline \texttt{[ \newline \{"@value": "Lei Complementar nº 123 de 14/12/2006", \newline "@language": "pt"\}, \newline \{"@value": "Complementary Law nº 123 of 2006-12-14", \newline "@language": "en"\} \newline]} \\
\midrule

\url{alternateName} \newline (\texttt{R64 has name}) & List of common or informal names for the legal \emph{Work} (\texttt{F52 Name}). Optional. & A simple list or a list of language-localized values: \newline \texttt{["Lei do Supersimples", "Lei Geral das Micro e Pequenas Empresas"]} \newline Or \newline \texttt{[ \newline \{"@value": "Lei do Supersimples", \newline "@language": "pt"\}, \newline \{"@value": "Law of Supersimple", \newline "@language": "en"\}, \newline \{"@value": "Lei Geral das Micro e Pequenas Empresas", \newline "@language": "pt"\}, \newline \{"@value": "General Law of Micro and Small Enterprises", \newline "@language": "en"\} \newline ]} \\
\midrule

\url{abstract} * & The official summary (ementa) of the legal \emph{Work}---as a component of the text, it is a characteristic of a specific F2 Expression. The value provided here is typically inferred from a representative or the most recent expression. Optional. & A simple value or a list of language-localized values: \newline "\texttt{Institui o Estatuto Nacional da Microempresa e da Empresa de Pequeno Porte; altera dispositivos das Leis nºs 8.212 e 8.213, ambas de 24 de julho de 1991, da Consolidação das Leis do Trabalho - CLT, aprovada...}" \newline Or \newline \texttt{[ \newline\{"@value": "Institui...", \newline "@language": "pt"\}, \newline \{"@value": "Establish...", \newline"@language": "en"\} \newline ]} \\
\midrule

\url{keywords} \newline (\texttt{P67 refers to}) & List of keywords (\texttt{E55 Type}) for indexing the legal \emph{Work}. Optional. & A simple list or a list of language-localized values: \newline \texttt{["Pequena Empresa", "Microempresa"]} \newline Or \newline \texttt{[ \newline \{"@value": "Pequena Empresa", \newline "@language": "pt"\}, \newline \{"@value": "Small Enterprise", \newline "@language": "en"\}, \newline \{"@value": "Microempresa", \newline "@language": "pt"\}, \newline \{"@value": "Micro enterprise", \newline "@language": "en"\}\newline ]} \\
\midrule

\url{about} \newline (\texttt{P67 refers to}) & List of classification themes (\texttt{E55 Type}) for the legal \emph{Work} \footnote{The set of possible values for themes of Brazilian legislation can be found at \url{https://normas.leg.br/conceito/?urn=urn:lex:br:federal:conceito:tema}. Last accessed on 2025-04-19.}. Optional. \newline The value of the \texttt{name} property can be provided, for display purposes, when the source doesn't have a structured Schema.org description for the object. & Hypothetical example: \newline \texttt{ \newline \{ "@type": "DefinedTerm", "@id": "\url{https://normas.leg.br/tema/?urn=urn:lex:br:federal:tema:direito.empresarial.e.economico}" \}\newline } \\
\midrule

\url{legislationDate} * & The date of enactment (signature), marking the creation of the legal \emph{Work} in YYYY-MM-DD format. \newline As presented above, the norm's F1 Work was created through an \texttt{F28 Expression Creation} event, which represents the production of the first official expression of the norm. The date itself is represented by an \texttt{E52 Time-Span}, linked to this event via (\texttt{P4 has time-span}). & "2006-12-14" \\
\midrule

\url{legislationPassedBy} * & The legislative body that passed the legal \emph{Work}. \newline According to its definition as the "legal author", this property maps to the legislative body that debated and approved the norm's content: the \texttt{F11 Corporate Body} representing the Parliament. In a high-fidelity LRMoo model, this would be the agent of a distinct "Approval Event" that precedes the "Enactment/Sanction Event" (normally carried out by the Executive). & \texttt{\{ "@type": "GovernmentOrganization", "@id": "\url{https://www.congressonacional.leg.br/}" \}} \\
\midrule

\url{spatialCoverage} & The geographical area of applicability. It can be, for example, the country, a state, a municipality\footnote{As the URI for Brazilian locations, the official registry of IBGE (Brazilian Institute of Geography and Statistics) can be used: \url{https://servicodados.ibge.gov.br/api/docs/localidades}. Last accessed on 2025-04-19.}. The geographical area of applicability is modeled as an \texttt{E53 Place}. While no direct property links an F1 Work to a Place, it's often inferred from the jurisdiction or the creating body's scope or from the the textual content of the legal norm. \newline The value of the \texttt{name} property can be provided, for display purposes, when the source doesn't have a structured Schema.org description for the object. & \texttt{\{ "@type": "Country", "@id": "\url{https://servicodados.ibge.gov.br/api/v1/localidades/paises/76}", "name": "Brasil (país)", "url": "\url{https://www.wikidata.org/wiki/Q155}", "address": \{ "@type": "PostalAddress", "addressCountry": "BR" \} \}} \\
\midrule

\url{legislationJurisdiction} * & This property identifies the political or administrative entity that holds the legal authority to promulgate the legislative instrument. The jurisdiction is the sovereign \texttt{F11 Corporate Body} (e.g., "União"). It is linked to the agent of the creation event (legislationPassedBy) via the property \texttt{P107 is part of}. \newline It pertains to the concept of sovereignty and competence, answering the question "who" is the source of the law's power, rather than "where" the law applies, which is addressed by \texttt{spatialCoverage}. \newline For example, a federal law could originate from the federal jurisdiction of the Union, but its spatialCoverage could be restricted to the specific territory of the Manaus Free Trade Zone ("Zona Franca de Manaus"). \newline For federal law, the jurisdiction is the federal union; for state law, it is the respective state etc. & \texttt{\{ "@type": "AdministrativeArea", "@id": "\url{https://www.wikidata.org/wiki/Q5440531}", "name": "Governo Federal do Brasil (União)"\}} \\
\midrule

\url{datePublished} * &  The publication date of the original version of the legal \emph{Work}, in YYYY-MM-DD format. \newline Temporal description of the norm's publication event. Describes the Publication Event of the containing Official Gazette edition. The date is an \texttt{E52 Time-Span} (\texttt{P4 has time-span}) of this aggregation event. & "2006-12-15" \\
\midrule

\url{publisher} * & The official publisher of the original version of the legal \emph{Work}. \newline Spatial (\texttt{locationPublished}) and Institutional (agent) description of the norm's publication event. The official publisher is the agent, an \texttt{F11 Corporate Body}, that carried out the Publication Event of the containing Official Gazette edition. & \texttt{\{ "@type": "GovernmentOrganization", "@id": "\url{https://www.in.gov.br/}" \}} \\
\midrule

\url{license} * & License for the publication of the original version of the legal \emph{Work}. Describes the publication terms of original version \texttt{F3 Manifestation}. LRMoo/CRM models this as a \texttt{E30 Right} associated with the publication event via \texttt{P30i has right}. & "\url{http://creativecommons.org/licenses/by/4.0/}" \\
\midrule

\url{legislationLegalForce} * & Indicates whether this legal \emph{Work} is currently in force.  It can be inferred by the validity period of the last F2 Expression of the F1 Work. Optional. \newline Should only be specified, with "NotInForce", when the norm is not currently in force. & "NotInForce" (Hypothetical example) \\
\midrule

\url{temporalCoverage} * & Period during which this legal \emph{Work} was in force. Optional. It describes the temporal extent during which some version of the legal Work was in force. This is often represented as the union of the validity periods of its successive F2 Expressions. This is modeled as the \texttt{E52 Time-Span} (\texttt{P4 has time-span}) of the F2 Expression itself. \newline 
According to project ELI\footnote{The European Legislation Identifier (ELI) standard, whose ontology was a key input for the \texttt{Schema.org/Legislation} vocabulary. For more details, see the official documentation \cite{eli-guide}.}: "in force range of the act, from the date it was set in force to the date it was repealed." \newline 
Should only be specified when: 
\begin{itemize}[noitemsep,topsep=0pt]
    \item this legal \emph{Work} is not currently in force (e.g., "2006-12-20/2010-01-05");
    \item the legal \emph{Work}'s effective date differs from its default start validity date (e.g., "2006-12-20/..").
\end{itemize}
May contain multiple non-continuous periods (e.g., \texttt{["2006-12-20/2010-01-05", "2012-05-05/.."]})  (International Standard ISO 8601). & "2006-12-14/2018-03-30" \\
\midrule

\url{legislationDateOfApplicability} * & The date at which the legal \emph{Work} becomes applicable. It describes the date of applicability of the last F2 Expression of the F1 Work. Optional. \newline Should only be specified when this date is different from the date of entry into force - for example, a legal \emph{Work} may come in force today, and state it will become applicable in 3 months. & "2007-03-14" (Hypothetical example) \\
\midrule

\url{sdPublisher} ** & Publisher of the structured data for this legal \emph{Work}. & \texttt{\{ "@type": "GovernmentOrganization", "@id": "\url{https://www.congressonacional.leg.br/}" \}} \\
\midrule

\url{sdDatePublished} ** & Date of publication of the structured data for this legal \emph{Work}, in YYYY-MM-DD format. & "2021-10-04" \\
\midrule

\url{sdLicense} ** & License for the publication of the structured data for this legal \emph{Work}. & "\url{http://creativecommons.org/licenses/by/4.0/}" \\
\bottomrule

\end{longtable}

* Properties marked with an asterisk represent metadata that are not intrinsic to the abstract F1 Work. Instead, they are convenient shortcuts semantically derived from other related entities, such as: (a) the norm's creation and publication events (e.g., legislationDate, datePublished, publisher); (b) its versioned textual content, the F2 Expressions (e.g., abstract, temporalCoverage); or (c) its physical or digital embodiment, the F3 Manifestations (e.g., license). In a simplified Work-centric description, these are included for ease of consumption. However, in a complete, multi-version knowledge graph (or in complete, multi-version JSON-LD, as described in a forthcoming paper), these properties would be omitted from the F1 Work's description and instead be explicitly modeled on their respective source entities.

** Properties marked with a double asterisk describe the provenance of the structured data record itself, not the legal norm. In CIDOC CRM, this is modeled as an E65 Creation event for the metadata record, with its own agent (E39 Actor), date (E52 Time-Span), and rights (E30 Right).

To illustrate how the properties detailed in Table \ref{tab:norm_properties} come together, Listing \ref{fig:jsonld_norm_example} presents a complete JSON-LD document\footnote{The JSON-LD examples can be tested using the Schema.org validator: \url{https://validator.schema.org/}. Last accessed on 2025-04-19.} for the legal \textit{Work} "Lei Complementar nº 123 de 14/12/2006". This example demonstrates a structured representation of this conceptual entity according to the Schema.org vocabulary. While some property values are hypothetical for didactic purposes, the structure reflects the proposed mapping.

The example also highlights three key properties that operationalize the Linked Data principles discussed previously, transforming the \textit{Work} entity into a discoverable and interconnected node on the web of data:
\begin{itemize}[leftmargin=*]
    \item \texttt{@context}: Maps the terms used in the document (e.g., "name", "legislationDate") to their full IRIs in the Schema.org vocabulary.
    \item \texttt{@type}: Explicitly declares the resource's type, in this case, \texttt{sdo:Legislation}.
    \item \texttt{@id}: Provides a unique, global identifier (URI) for the resource being described, making it linkable from anywhere on the web.
\end{itemize}

\begin{lstlisting}[
    caption={A json-LD with general Legislation properties of the Brazilian legal \textit{Work} Complementary Law No. 123 of 2006-12-14.},
    label={fig:jsonld_norm_example},
    breaklines=true,
    float=false,
    basicstyle=\ttfamily\tiny,
    escapechar=| % Define '|' as the escape character
]
{
  "@context": "http://schema.org/",
  "@type": "Legislation",
  "@id": "https://normas.leg.br/?urn=urn:lex:br:federal:lei.complementar:2006-12-14;123",
  "legislationType": {
    "@type": "CategoryCode",
    "@id": "https://normas.leg.br/?urn=urn:lex:br:federal:tipo-norma:lei.complementar"
  },
  "legislationIdentifier": "urn:lex:br:federal:lei.complementar:2006-12-14;123",
  "sameAs": [
    "http://www.lexml.gov.br/urn/urn:lex:br:federal:lei.complementar:2006-12-14;123",
    "http://legis.senado.leg.br/legislacao/DetalhaSigen.action?id=572878"
  ],
  "inLanguage": "pt",
  "name": "Lei Complementar n|\textordmasculine{}| 123 de 14/12/2006",
  "alternateName": [
    "LCP-123-2006-12-14",
    "Lei do Supersimples",
    "Lei Geral das Micro e Pequenas Empresas"
  ],
  "abstract": "Institui o Estatuto Nacional da Microempresa e da Empresa de Pequeno Porte; altera dispositivos ...",
  "about": { 
    "@type": "DefinedTerm", 
    "@id": "https://normas.leg.br/tema/?urn=urn:lex:br:federal:tema:direito.empresarial.e.economico" 
  },  
  "keywords": [
    "Microempresa",
    "Pequena Empresa"
  ],
  "legislationDate": "2006-12-14",
  "legislationPassedBy": {
    "@type": "GovernmentOrganization",
    "@id": "https://www.congressonacional.leg.br/"
  },
  "legislationJurisdiction": {
    "@type": "AdministrativeArea",
    "@id": "https://www.wikidata.org/wiki/Q5440531",
    "name": "Governo Federal do Brasil (Uniao)"
  },
  "spatialCoverage": {
    "@type": "Country",
    "@id": "https://servicodados.ibge.gov.br/api/v1/localidades/paises/76",
    "name": "Brasil",
    "url": "https://www.wikidata.org/wiki/Q155",
    "address": {
      "@type": "PostalAddress",
      "addressCountry": "BR"
    }
  },
  "datePublished": "2006-12-15",
  "license": "https://creativecommons.org/licenses/by/4.0/",
  "publisher": {
    "@type": "GovernmentOrganization",
    "@id": "https://www.in.gov.br/"
  },
  "sdDatePublished": "2024-10-04",
  "sdLicense": "https://creativecommons.org/licenses/by/4.0/",
  "sdPublisher": {
    "@type": "GovernmentOrganization",
    "@id": "https://www.congressonacional.leg.br/"
  }
}
\end{lstlisting}

\section{Conclusion and Future Work}

This paper, the first in a foundational series, detailed the essential first step in translating a formal, LRMoo-based model of legal norms \cite{demartim2025temporal} to the web. We presented a detailed mapping of the abstract F1 Work—the conceptual and persistent identity of a legal act—to the interoperable sdo:Legislation vocabulary. By providing a practical and standardized method for representing this stable identity as a URI-addressable entity, this work establishes the essential anchor for modeling the complete lifecycle of legal information.

This approach is critical for the future of Legal AI, as it provides a verifiable foundation that contrasts with the purely probabilistic nature of large language models. The key contributions of this work are:
\begin{itemize}
\item \textbf{A detailed property-by-property mapping} of the LRMoo \texttt{F1 Work} entity to \texttt{sdo:Legislation}, validated with a concrete JSON-LD implementation for the Brazilian legal system.
\item \textbf{An interoperable foundation for realizing the full, LRMoo-based lifecycle model of legal norms on the web}, which in turn enables the development of deterministic, temporally-aware Legal Knowledge Graphs.
\end{itemize}

By encoding the core identity of legal information in a standardized, machine-readable format, this work establishes a verifiable "ground truth" essential for advancing beyond probabilistic models toward deterministic legal information retrieval and trustworthy decision support systems.

For future work, this well-defined F1 Work entity serves as the cornerstone for the subsequent parts of this series. As outlined in our foundational ontology \cite{demartim2025temporal}, the immediate next steps involve the mapping of the versioning and structural layers. Part II will detail the representation of F2 Expressions (temporal and language versions), while subsequent work will address the norm's internal hierarchy via Component Works. Completing this multi-layered model will yield a comprehensive Legal Knowledge Graph, enabling advanced, structure-aware applications that can deterministically reconstruct the law as it existed at any point in time.

\bibliographystyle{ios1} 
\bibliography{referencias}

\end{document}